# Polarization Multiplexed Metalens Array Optical Chip for High-Performance LWIR Polarimetric Camera

Shichuan Wang[1†], Tie Hu[1†], Zihan Mei[1], Xuancheng Peng[1], Bing Yan[2], Wenhong Zhou[2], Ming Zhao[1] and Zhenyu Yang[1‡]

[1]School of Optical and Electronic Information, Huazhong University of Science and Technology, Wuhan 430074, China.

[2]Wuhan Global Sensor Technology Co., Ltd., Wuhan 430205, China.

**Abstract**

Compared with traditional infrared thermal imaging, polarimetric imaging provides additional polarization information, which effectively enhances object contours and image contrast, with broad application in both military and civilian domains. However, the traditional long-wave infrared polarimetric camera suffers from severe thermal noise, low sensitivity and limited detection accuracy. To address the aforementioned problems, a novel cooled LWIR polarimetric camera based on an achromatic polarization multiplexed germanium-based metalens array optical chip is reported in this paper, enabling high-precision division of focal plane linearly polarimetric imaging. The proposed system demonstrates high-precision linearly polarimetric imaging, with the metalens array achieving an average transmittance of 84.7% across the 8.4~11.6μm band and a polarization extinction ratio exceeding 10. The metasurface-based camera attains an average polarization reconstruction error below 0.981%, markedly surpassing state-of-the-art commercial LWIR polarimetric systems. Additionally, the new camera presents excellent polarimetric imaging capability for complex scenes. To the best of our knowledge, this represents the world's first LWIR polarimetric camera utilizing the metasurface optical chip with performance superior to commercial cameras, promoting the practical development of metasurface-integrated devices.

**Introduction**

The long-wave infrared (LWIR) spectral band, spanning 8~14μm, exhibits minimal atmospheric absorption and ubiquitous thermal radiation from objects with temperatures above absolute zero, making it particularly well-suited for passive thermal imaging applications across diverse domains[1–3]. Compared with the traditional LWIR detectors which are limited to the intensity information of the incident light, the polarization characteristics of the target can provide richer information[4–6]. LWIR polarimetric imaging technology demonstrates significant potential in anti-jamming target recognition and artifact identification within complex environment [7,8]. Due to the characteristics of compact structures, high real-time performance and easy integration, division of focal plane (DoFP) polarimetric detectors have been widely studied[9–11]. Polaris Sensor Technology successfully developed the first uncooled LWIR polarimetric camera, Pyxis®LWIR 640, and systematically researched the application scenarios and advantages of long-wave infrared polarimetric imaging[12]. However, owing to the design scheme of micro polarizer arrays[13,14], its theoretical efficiency cannot exceed efficiency ceilings of 50%, and the ohmic loss in the metal micro polarizers further reduces the efficiency of the device and weakens the low-light detection capability. Moreover, the absence of microlens arrays for focusing exacerbates inter-pixel crosstalk, substantially degrading imaging fidelity.

Metasurface utilizes meta-atoms to simultaneously and independently modulate the phase and polarization state of incident light at subwavelength spatial scales, offering a pathway to enhance the functionality and integration of optical systems[15]. Yuxi Wang et al. proposed an optical multi-parameter detector based on near-infrared silicon-based metalens array[16], where each super unit consists of four sub-metalenses designed for 0° linearly polarized light (XLP), 45° linearly polarized light (ALP), 90° linearly polarized light (YLP), and left circularly polarized light (LCP) detection. Compared with the polarimetric camera based on micro polarizer array, the metalens array achieves both polarization filtering and beam convergence without requiring complex alignment and integration processes. The metasurface-based polarimetric camera shows higher efficiency, lower signal crosstalk, and superior polarization detection accuracy. Similarly, some scholars have proposed polarization sensitive metalens designs with spatial ring-band splitting[17] and spatial interpolation multiplexing[18]. Although the light convergence effect of the aforementioned polarization sensitive metalens increases the photon collection efficiency, the polarization-filtered polarization state modulation effect introduces optical energy loss. To overcome this drawback, Ehsan Arbabi et al. developed a visible light polarimeter based on a polarization multiplexed metalens array with near-100% energy utilization [19]. Nevertheless, the severe radiation interference in the LWIR band restricts most of the current metasurface-based polarimeters to visible and near-infrared bands, leaving a critical research gap.

In this work, a novel cooled LWIR polarimetric camera employing an achromatic polarization multiplexed germanium-based (Ge-based) metalens array is proposed and validated. The core chip of the polarimetric camera consists of a Ge-based metalens array optical chip and an InAs/GaSb type II superlattice infrared focal plane chip[20–22], achieving DoFP polarimetric imaging. Each super unit contains an XLP-YLP and an ALP-BLP (135° linearly polarized light) polarization multiplexed metalens. Based on the principles of discrete multi-wavelength achromatic aberration[23–25] and polarization multiplexing[26,27], the metalens array optical chip simultaneously accomplishes polarization beam splitting and light convergence, with an average transmittance higher than 84.7% (in the wavelength band of 8.4~11.6μm) surpassing the limit (50%) of polarizers, and an average polarization extinction ratio higher than 10. Under broadband blackbody illumination, the mean absolute value of the polarization angle (AOP) error of the new infrared polarimetric camera is 0.21°, the mean absolute value of the degree of linearly polarization (DOLP) error is 0.009, and the average polarization reconstruction error of the super unit is 0.981%. Under the same test conditions, the mean absolute value of AOP error of commercial LWIR polarimetric cameras is 0.53°, the mean absolute value of DOLP error is 0.015, and the average polarization reconstruction error is 2.700%. The cooled LWIR polarimetric camera incorporating the achromatic Ge-based metalens array demonstrates enhanced polarization detection accuracy.

## Results
### Concept and design

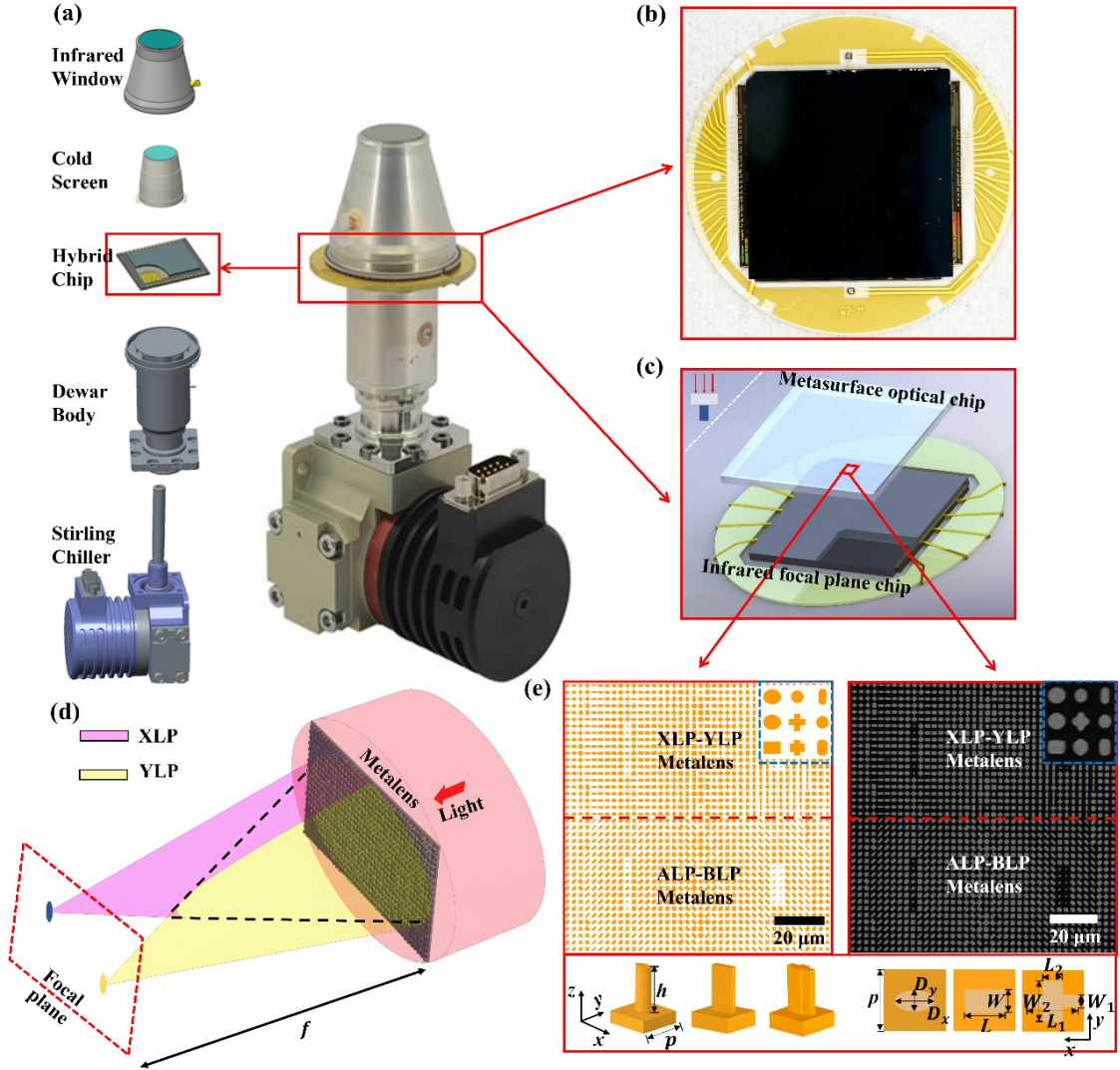

**Fig. 1 Principle of the metasurface-based LWIR polarimetric camera. a** Physical drawing of the metasurface-based LWIR polarimetric camera, with the camera structure schematic on the left. **b** Physical drawing of the core chip. **c** Schematic diagram of the core chip, with the light incidence mode in the upper left corner. d Schematic principle of the XLP-YLP polarization multiplexed metalens, and $f = 42\mu m$. **e** Layout of the super unit design (left) and the SEM image (right), and the meta-atomic structure used for the design at the bottom. The geometric parameters are as follows: $h = 12\mu m$, $p = 2.5\mu m$. $D_x$, $D_y$, $W$, $L$, $W_1$, $W_2$, $L_1$, and $L_2$ are geometric variables.

Fig. 1a displays the physical diagram of the metasurface-based LWIR polarimetric camera, comprising an infrared window, a cold screen, a hybrid chip, a Dewar body, and a Stirling cooler. The core chip consists of a Ge-based metalens array optical chip tightly pasted with an InAs/GaSb type II superlattice infrared focal plane chip, as shown in Fig. 1b. Fig. 1c demonstrates the schematic diagram of the core chip, showing the metasurface optical chip aligned with the focal plane chip for integration.

The polarization multiplexed metalens design leverages the capability of the subwavelength meta-atoms to independently regulate the electromagnetic response of the orthogonally polarization components of the incident light. Through reasonable phase design, the incident light is focused into spatially separated off-axis focal points corresponding to orthogonal polarization states, enabling a theoretical near-100% energy utilization, as shown in Fig. 1d. According to the generalized law of refraction[28], the wavefront of the ideal focusing beam satisfies the hyperbolic phase distribution, and the phase profile of the achromatic XLP-YLP polarization multiplexing metalens can be expressed as:

$$\varphi_{XLP}(x,y,\lambda) = \frac{2\pi}{\lambda}\left(f - \sqrt{(x+d_x)^2 + (y-d_y)^2 + f^2}\right) + C_{XLP}(\lambda) \quad (1)$$

$$\varphi_{YLP}(x,y,\lambda) = \frac{2\pi}{\lambda}\left(f - \sqrt{(x-d_x)^2 + (y-d_y)^2 + f^2}\right) + C_{YLP}(\lambda) \quad (1)$$

where $\varphi_{XLP}$ and $\varphi_{YLP}$ are the ideal phase distributions of the polarimetric multiplexed off-axis focusing metalens corresponding to XLP and YLP light incidence, respectively. $\lambda$ denotes the design wavelength, which is in the range of 8.4~11.6μm. $(x, y)$ are the meta-atomic coordinates, and $(-d_x, d_y, f)$ and $(d_x, d_y, f)$ denote the corresponding focal coordinates of the metalens under XLP and YLP light incidence, with the center of the metalens as the XOY coordinate origin. $d_x = 25\mu m$, $d_y = 0$, $f = 42\mu m$ represents the design focal lengths, and $C_{XLP}(\lambda)$ and $C_{YLP}(\lambda)$ are the wavelength-dependent phase bias terms.

The metasurface optical chip is designed using low absorption loss Ge ($\lambda = 10\mu m$: $n = 4.004$, $k = 0.000$) wafers in the LWIR band, with a meta-atomic period of $p = 2.5\mu m$ and a dielectric column height of $h = 12\mu m$. The broadband design of 8.4~11.6μm puts a requirement on the bulk of the structural library, for which we chose three different geometrical types of structure, with the dielectric column cross-sectional shape of elliptic, rectangular, and cross, respectively. The optical response of the meta-atom is mainly affected by the structure filling factor, so the finite time domain difference (FDTD) algorithm is used for the size sweeping simulation[29], and the simulation results are shown in Supplementary Information S1. The broadband design adopts a discrete multi-wavelength design, and the wavelength sampling points are set to be 8.4~11.6μm, with the spacing of 0.4μm. The phase constant term C(λ) at each wavelength is first optimized to match the structure library. Subsequently, the meta-atom array exhibiting the minimal average phase error and maximum transmittance is selected according to the phase profile described in Eqs. (1) and (2). The ALP-BLP polarization multiplexed metalens is obtained by rotating the meta-atoms by 45° on the basis of the XLP-YLP polarization multiplexed metalens. Assembling achromatic XLP-YLP and ALP-BLP polarization multiplexed metalenses forms a 100μm×100μm super unit, consisting of 40×40 meta-atoms. 160×128 super units are spliced to form a metasurface optical chip, with the dimension of 16mm×12.8mm. Fig. 1e displays both the design of the super unit and the SEM image of processed structures.

**Simulation of polarization multiplexed metalens**

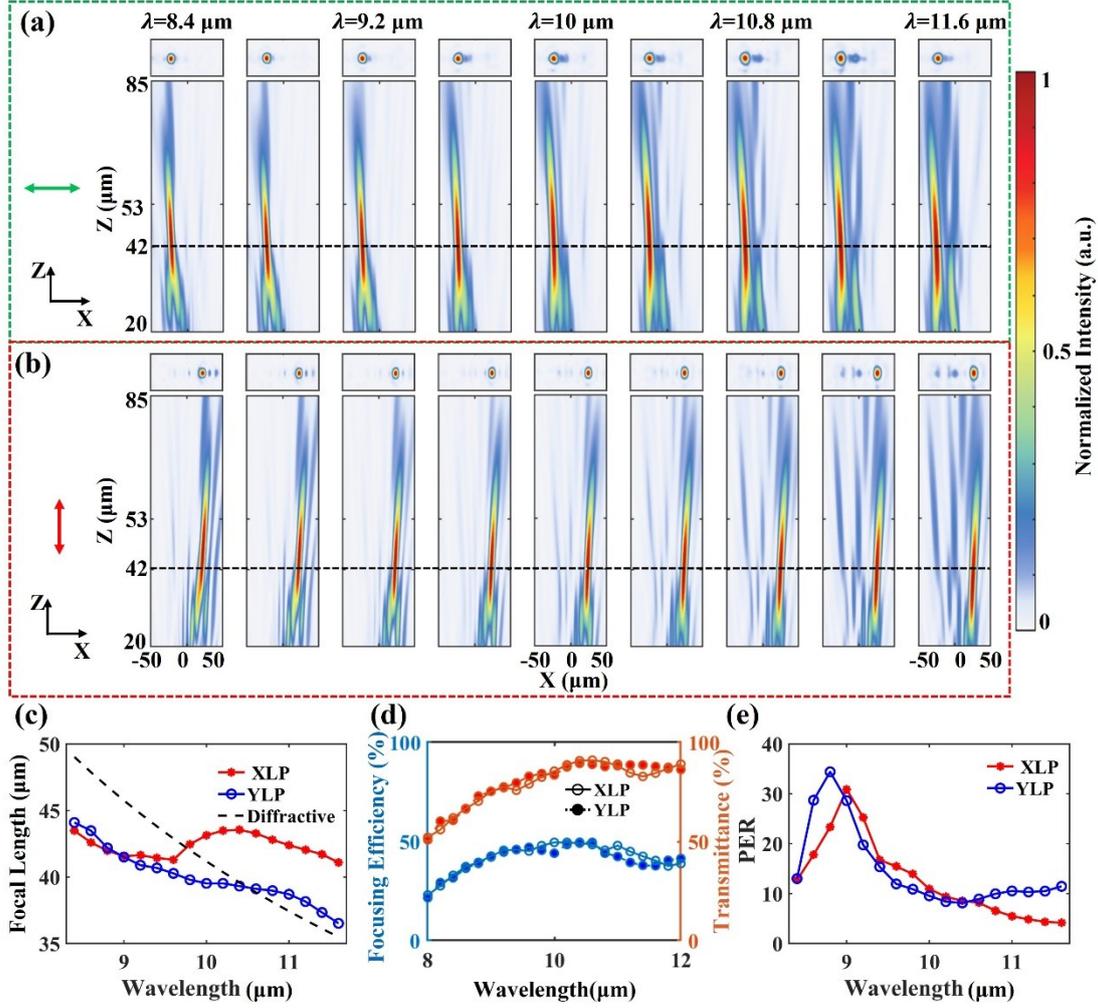

**Fig. 2 Polarization response of XLP-YLP polarization multiplexed metalens.** Normalized light field distribution with XLP light incidence (**a**) and YLP light incidence (**b**). **c** Simulated focal length. **d** Transmittance and absolute focusing efficiency. **e** Polarization extinction ratio.

Fig. 2a and Fig. 2b show the XZ longitudinal normalized light field distributions and the XY transverse normalized light field distributions at the focal plane of the XLP-YLP polarization multiplexed metalens under the incidence of XLP and YLP light, respectively. The average focal length of the metalens in the broadband range of 8.4~11.6μm is approximately 42μm, and the metalens focuses off-axis at the X-coordinate values of -25μm and 25μm under XLP and YLP illumination, respectively, in strong agreement with design. The elliptical focal spot stems from differing apertures of orthogonal axes in the polarization multiplexed metalens. Fig. 2c demonstrates the simulated focal lengths of the broadband achromatic metalens, ranging from 41.3 to 43.6μm with a maximum focal length shift less than 3.0% under XLP illumination, and from 36.5 to 44.1μm with a maximum focal length shift less than 9.8% under YLP illumination. The focal length range of the diffractive metalens is 35.5~49.2μm with a maximum focal length shift of 17.1%. The broadband achromatic metalens exhibits excellent achromatic properties. The focal length shift is defined as the ratio of the absolute difference between the focal length and the average focal length at the corresponding wavelength to the average focal length. Fig. 2d demonstrates the absolute focusing efficiency and transmittance versus wavelength for the broadband achromatic metalens under XLP and YLP light incidence, respectively. The transmittance of the metalens remains above 50% with an average value of 78.99% (XLP) as well as 79.75% (YLP). Absolute focusing efficiencies are 41.98% (XLP) and 40.86% (YLP), respectively. The energy loss mainly originates from the diffraction loss. The absolute focusing efficiency is defined as the ratio of the total power in the focal plane within the range of 3 times FWHM (full width at half maximum) to the total incident power. Fig. 2e shows the polarization extinction ratio of the XLP-YLP polarization multiplexed metalens, which is defined as the ratio of the absolute focusing efficiency of the corresponding polarization to the focusing efficiency of the orthogonal polarization. When illuminated with XLP light, the polarization extinction ratio of the metalens achieves 30.9 at $\lambda = 9\mu m$. When illuminated with YLP light, the polarization extinction ratios consistently above 8.1 in the whole wavelength range, reaching 34.4 at $\lambda = 8.8\mu m$. The achromatic polarization multiplexed metalens maintains robust polarization-splitting performance throughout the broad bandwidth. The ALP-BLP polarization multiplexed metalens exhibits comparable performance, with detailed characterization data provided in Supplementary Information S2.

**Polarization detection performance of metasurface optical chip**

To minimize specular reflection loss, we employ germanium (Ge) wafers with a LWIR transmittance-enhancing coating deposited on one side. Fig. 3b shows the transmittance spectra of the metasurface chip measured by a Fourier infrared spectrometer (Thermo Scientific, Nicolet iS50R). Compared with the conventional single-side coated Ge wafer (average transmittance 67.8%), the average transmittance of the metasurface optical chip is 84.7% in the band of 8.4-11.6μm, matching the simulation results. It indicates that the meta-atom layer acts as a low-refractive-index matching layer, thus providing a transmittance-enhancing effect. The optical chip performance was characterized by the optical setup shown in Fig. 3a, with polarizer 1 regulating the light intensity, polarizer 2 adjusting the polarization state of the light, and the home-made microscope system imaging the metasurface back-focal plane spot array to the CCD image sensor. To characterize the broad-spectrum properties of the chip, two lasers with central wavelengths of 9.3μm and 10.6μm and bandwidths of about 50nm were used sequentially for the experiments. Fig. 3c demonstrates the normalized light intensity distribution in the focal plane of a super unit under linearly polarized light incidence. When subjected to XLP, YLP, ALP and BLP illumination, each super unit exhibits maximum focused spot intensity in its designated polarization region while minimal in the orthogonally polarized counterpart. Fig. 3d and f show the polarization response curves of the metasurface at 9.3μm and 10.6μm wavelengths. The response curves corresponding to the different focusing regions follow Malus's law, and the peak responses correspond to the polarization angles that differ by 45° in turn, which is consistent with the simulation results. The focusing regions are taken to be in the range of 3 times FWHM in the focal plane. The polarization extinction ratios at 9.3μm and 10.6μm are 13.5 (XLP), 19.9 (ALP), 19.4 (YLP), 14.2 (BLP), and 53.4 (XLP), 21.8 (ALP), 14.9 (YLP), 11 (BLP), respectively, confirming the effective polarization detection capability.

To characterize the polarization detection capability of the metasurface, we perform polarization reconstruction by focusing spots. The polarization information of the signal light can be represented by a Stokes vector[30]. Since the metalens designed is only sensitive to the linear polarization information, the three-dimensional Stokes vector is used, and its relationship with the light intensity can be expressed as $S = M \cdot I$, i.e:

$$\begin{bmatrix} S_0 \\ S_1 \\ S_2 \end{bmatrix} = \begin{bmatrix} M_{11} & M_{12} & M_{13} & M_{14} \\ M_{21} & M_{22} & M_{23} & M_{24} \\ M_{31} & M_{32} & M_{33} & M_{34} \end{bmatrix} \cdot \begin{bmatrix} I_{XLP} \\ I_{YLP} \\ I_{ALP} \\ I_{BLP} \end{bmatrix} \qquad (3)$$

where $I_{XLP}$, $I_{YLP}$, $I_{ALP}$, $I_{BLP}$ represent the light intensity received in the focusing region of the four linear polarization, $S_0$ represents the intensity of the incident light, $S_1$ reflects the intensity difference between the XLP and YLP components of the incident light, and $S_2$ reflects the intensity difference between the ALP and BLP components of the incident light. The reconstruction matrix $\boldsymbol{M}$ is solved inversely using six standard linear polarization states, and then the Stokes vector of the incident light is solved based on Eq. (3). Fig. 3e and Fig. 3g show the Poincaré sphere sections of the super unit at 9.3μm and 10.6μm, respectively, with average polarization reconstruction errors of 3.5% and 3.4%. Detailed polarization state reconstruction data are shown in Supplementary Information S6. The polarization reconstruction error is defined as the Euclidean distance of the reconstructed value from the theoretical value in the Poincaré sphere section as a percentage of unit length.

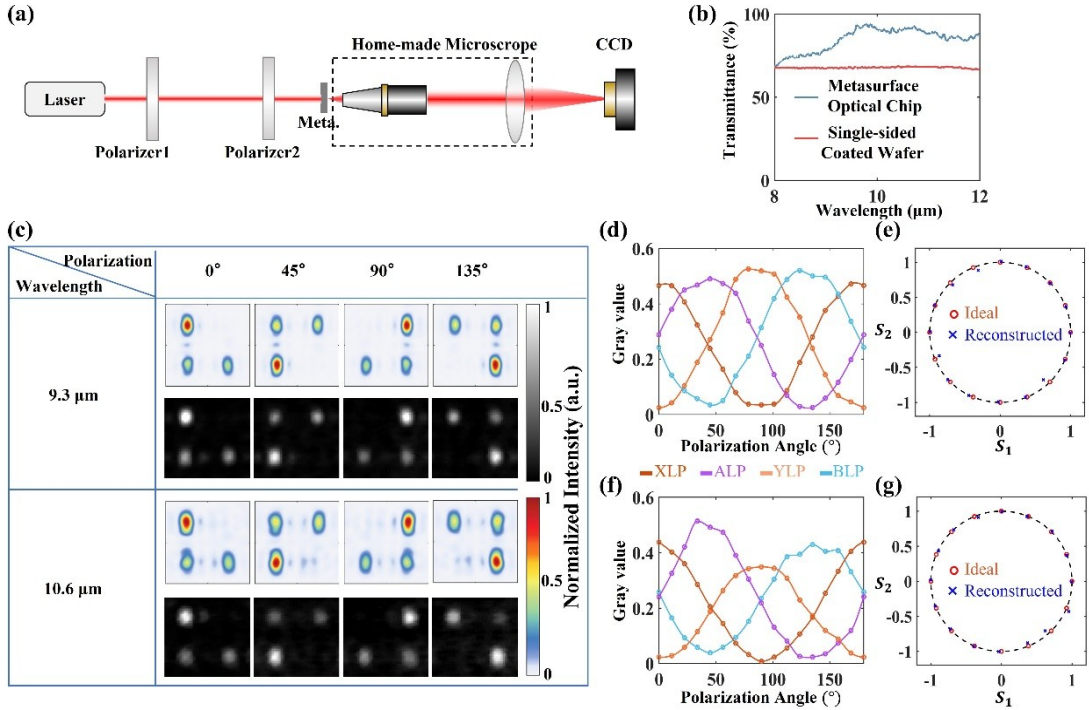

**Fig. 3 Polarization detection results of the metasurface optical chip. a** Polarization detection test optical setup. **b** Transmittance spectral line test results. **c** Focal plane normalized light intensity distributions of the polarization-detection super unit at 9.3μm and 10.6μm wavelengths for XLP, YLP, ALP, and BLP incidence, respectively. **d, f** Polarization response curves at 9.3μm and 10.6μm wavelengths. **e, g** Poincaré sphere sections of the metasurface at 9.3μm and 10.6μm wavelengths, with the dashed line indicating the unit circle.

**Metasurface-based LWIR polarimetric camera and characterization**

The core chip of the new LWIR polarimetric camera consists of a Ge-based metasurface optical chip and an InAs/GaSb type II superlattice infrared focal plane chip. The optical chip and the focal plane chip are integrated by flip-soldering technology[31]. To benchmark the performance of the novel LWIR polarimetric camera, we take a commercial uncooled LWIR polarimetric camera (North Guangwei Technology, GW5.0 PL0304X2A) for comparison. A thermal blackbody source couples with a wire-gird polarizer to generate broadband linearly polarized illumination. For each super unit, the incident light polarization state is reconstructed by combining Eq. (3) and error analysis is performed.

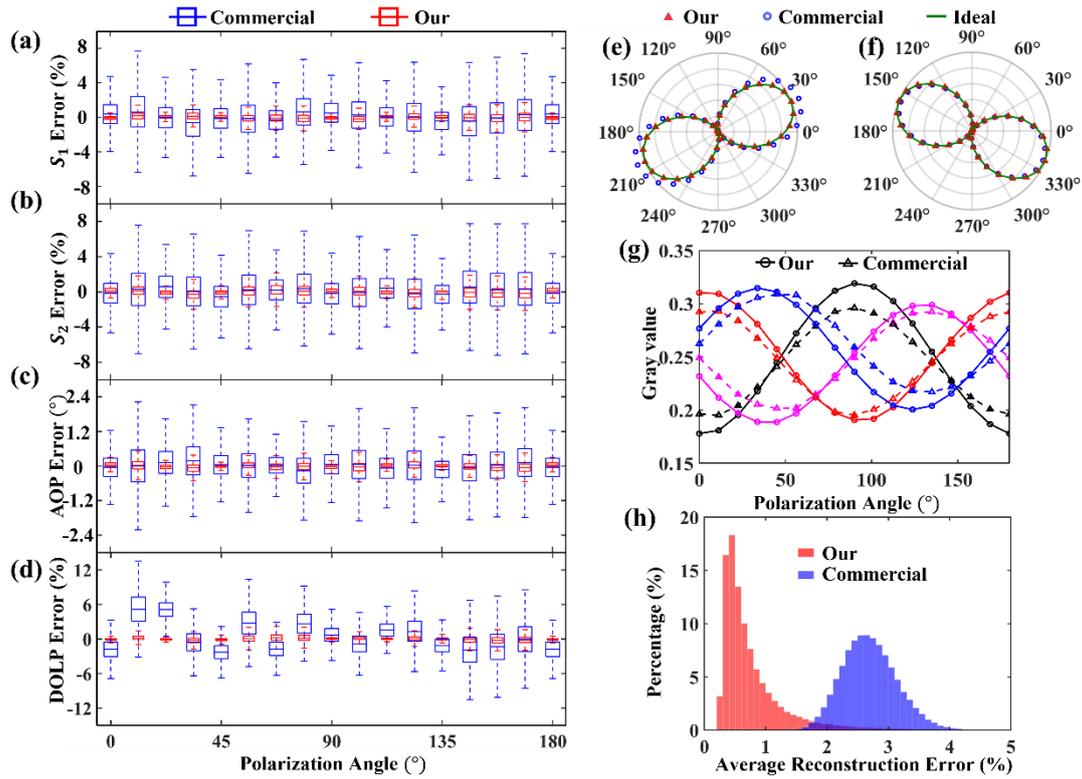

**Fig. 4 Comparison of polarimetric detection results between the new LWIR polarimetric camera based on the metasurface optical chip and the commercial LWIR polarimetric camera. a-d** Box line comparison plots of $S_1$, $S_2$, AOP and DOLP measurement errors of the two cameras, with the inner and outer boundaries of the box line plots representing the 50% and 75% occupation lines, respectively. **e, f** Polarization plots of two linearly polarization states. **g** Comparison of the polarization response curves of the two cameras. The triangular and circular line plots represent the test data of the commercially available and the metasurface-based LWIR polarimetric camera, respectively. **h** Comparison of the histograms of the average reconstruction error distributions of the two cameras.

Fig. 4a~d show the statistical results of S1, S2, AOP, and DOLP reconstruction errors of the two cameras under 16 broadband linearly polarized light incidence, respectively. The AOP reconstruction error is defined as the difference between the measured value and the ideal value, the S1 and S2 reconstruction errors are defined as the ratio of the difference value to S0, and the DOLP reconstruction error is defined as the ratio of the difference value to the ideal value. The metasurface-based LWIR polarimetric camera demonstrates superior accuracy, with >50% of super units achieving reconstruction errors within ±1% for $S_1/S_2$ and <0.18° for AOP. In contrast, >50% of super units in the commercial camera exhibit larger errors (±2% for $S_1/S_2$ and >0.66° for AOP). Moreover, the DOLP error of the novel LWIR polarimetric camera is significantly lower than that of the commercial camera. Fig. 4e and Fig. 4f show the polarization plots of two linear polarization states (polarization angles of 22.5° and -22.5°), indicating higher reconstruction accuracy of the metasurface-based LWIR camera. Fig. 4g demonstrates the normalized polarimetric response curves of the two cameras, following Malus' law. But the novel LWIR polarimetric camera offers more accurate peaks, troughs and greater contrast, which results in higher polarimetric sensitivity. The statistical distribution of the average polarization reconstruction error is shown in Fig. 4h. The average reconstruction error of the commercial LWIR polarimetric camera has a statistical mean of 2.700%, while that of the novel LWIR polarimetric camera achieves 0.981%.

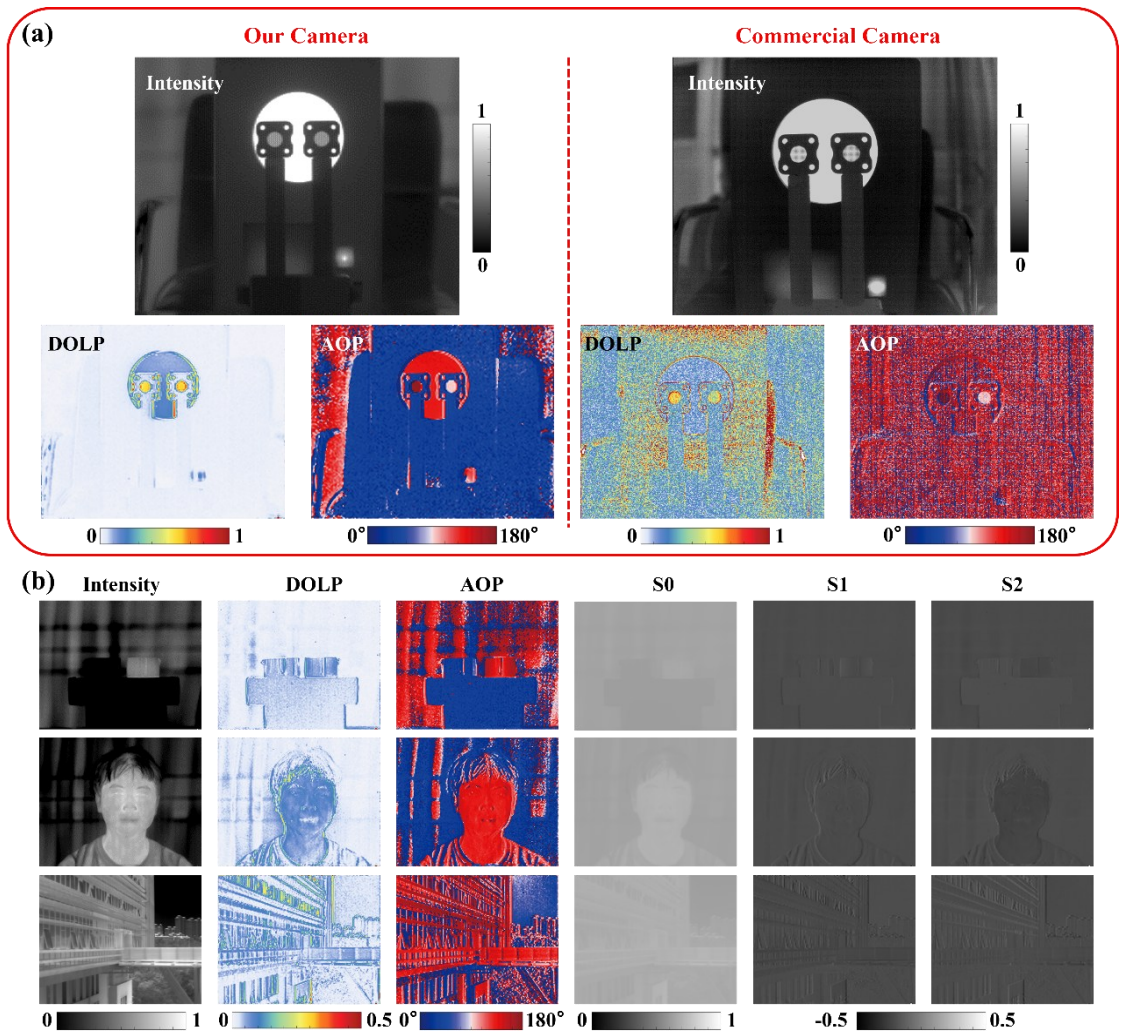

**Fig. 5 Polarimetric imaging results. a** Comparison of polarizer imaging results, with the novel LWIR polarimetric camera on the left and the commercial LWIR polarimetric camera on the right. **b** Complex scene imaging results demonstrated by the novel LWIR polarimetric camera. The targets are: glass cups with different temperatures, portraits and buildings.

Fig. 5a demonstrates a comparison of the imaging performance between our proposed new metasurface-based LWIR polarimetric camera and the conventional commercial camera with respect to polarizers of different polarization angles. The metasurface-based camera exhibits lower noise level and higher object contrast in the intensity map as well as DOLP image and AOP image, which are attributed to higher energy utilization and lower thermal noise. Due to the severe background radiation in the LWIR environment, the intensity figures are processed with linear contrast stretching to highlight the imaging target. The actual polarization angles of the left and right polarizers are 0° (horizontal) and 90° (vertical), and the average detection angles of the proposed metasurface-based camera and the commercial camera in the polarizer region are -2.4°/92.5° and -2.6°/94.2°, respectively. The DOLP in the polarizer region registers below unity (DOLP <1), owing to ambient thermal radiation interference during propagation. Fig.5b shows the imaging results of the proposed camera for complex scenes, containing imaging objects such as glass cups with different temperatures, portraits and buildings, respectively. While the low contrasts of S0, S1, and S2 images indicate that the LWIR radiations in the natural scene are disordered and seriously interfered, the DOLP and AOP images are able to identify the edges of the objects clearly and improve the contrasts of the objects significantly, confirming superior detection accuracy.

## Discussion

In this work, we have developed, to the best of our knowledge, the world's first LWIR polarimetric camera based on a polarization multiplexed achromatic metalens array optical chip. According to the discrete multi-wavelength achromatic theory, the Ge-based metalens array optical chip exhibits outstanding achromatic performance. The simulated average focal length in the 8.4~11.6μm wavelength band is approximately 42μm, with a maximum focal length shift of 9.8%. Experimentally, the metalens optical chip exhibits an average transmittance of 84.7% exceeding the limit (50%) of polarizers, and an average polarization extinction ration higher than 10 in LWIR band. The statistical mean value of the broadband radiation polarization reconstruction error of the proposed LWIR polarimetric camera is 0.981%, demonstrating high-precision polarimetric imaging performance. Collectively, the developed LWIR polarimetric camera shows lower signal crosstalk, higher efficiency and accuracy compared with the commercial LWIR polarimetric camera based on the micro polarizer array.

The development reported here bridges a critical gap in the application of metasurface technologies within the LWIR regime, representing a significant advancement in the practical implementation of flat optics. Polarimetric imaging technology based on the metasurface holds significant potential for applications in the fields of material identification[32], remote sensing[33], medical diagnosis[34] and astronomical observation[35]. With ongoing progress in multilayer meta-atom design methods[36,37], end-to-end inverse design methods[38], global optimization design methods[39–41], and machine learning design methods[42–44], the energy efficiency as well as the detection accuracy of LWIR polarization probing metasurface is expected to achieve further improvements.

## Material and methods

### Fabrication

We procured 500-μm-thick double-side-polished 2-inch Ge wafers from Shanghai Muhong Electronic Technology. To minimize the energy loss caused by specular reflection, we used chemical vapour deposition to carry out unilateral ZnS transmittance enhancement film evaporation, which can enhance the transmittance rate in the range of 8~12μm wide bandwidth.

For the metasurface fabrication process, a 5"×5"×0.09" blank lithography plate with a low reflectivity chromium layer (Telic Corporation) was used for the preparation of the lithography mask. Subsequently, contaminants such as metal ions, adsorbed particulates and organic contaminants, residual adsorbates, and natural oxidation layers were eliminated through organic and inorganic cleaning processes from the surfaces of germanium wafers, rendering the surfaces hydrophobic. After the surface preparation, an approximately 800-nm-thick layer of Beijing Kehua KMP 7600 photoresist was uniformly spin-coated on one side of the germanium wafer. This was followed by a soft baking process to ensure proper photoresist curing. The samples were exposed via UV lithography using a Nikon-NSR-2205i11D stepper photolithography machine and developed in tetramethylammonium hydroxide (TMAH) solvent. Subsequently, the exposed samples were hard baked with a solid film. For germanium wafers etching, a patterned photoresist layer served as the etching mask in an inductively coupled plasma system (Oxford Plasmalab, System100-ICP-180) using the developed deep etch Bosch process. The patterned photoresist residue was removed by immersion in N-methylpyrrolidone (NMP). This is followed by further removal of residual photoresist and minor contaminants using an oxygen plasma debonder.

We performed SEM measurements of the horizontal dimensions and etching depths of the meta-atoms, and the average processing errors were ±3.9% and ±2%, respectively, indicating high fabrication accuracy, and the detailed data are shown in Appendix S4.

**Integration**

In order to ensure precise alignment between the optical chip and the infrared focal plane chip, an alignment vernier is designed as shown in Fig. S9b, which is combined with flip chip technology and 3D integrated packaging technology to achieve high-precision integration. The working principle of the high-precision flip chip technology is shown in Fig. S9a, and the core is as follows: the infrared focal plane chip with the alignment vernier and the metalens array optical chip are aligned under the microscope, and the optical graphic is blended to form an overlapping image on the display to complete the alignment. Fig. S9c shows the structure diagram of the core chip of the LWIR polarimetric camera after integrated packaging.

**Acknowledgements**

This work was supported by Department of Science and Technology, Hubei Provincial People's Government (2021BAA003).

**Author details**

[1]Nanophotonics Laboratory, School of Optical and Electronic Information, Huazhong University of Science and Technology, Wuhan 430074, China.
[2]Wuhan Global Sensor Technology Co., Ltd., Wuhan 430205, China.

**Author contributions**

S. C. Wang and T. Hu provided the initial scheme for polarization detection, finished the simulations, and processed the data; S. C. Wang and Z. H. Mei completed the testing of the metasurface optical chip; X. C. Peng and B. Yan tested the imaging performance of the new camera and compared it with a commercial camera; W. H. Zhou and B. Yan tested the camera's focal plane chip and took charge of the core chip integration. M. Zhao, and Z. Y. Yang supervised the work and the manuscript writing. All authors discussed the results. S. C. Wang and T. Hu wrote the first draft of the manuscript, which was refined by contributions from all authors.

**Conflict of interest**

The authors declare no competing interests.

**Data availability**

The data that support the findings of this study are available from the corresponding author upon reasonable request.